\documentclass[12pt]{iopart}

\usepackage{microtype}
\usepackage{graphicx} %use graph format
\usepackage{epstopdf}
\usepackage{subfigure}
\usepackage{color, xcolor}
\usepackage{ulem}
\usepackage{soul}
\usepackage{hyperref}
\usepackage{cite}
\soulregister{\ref}7 %
\usepackage{color, xcolor}
\clearpage

\usepackage{lineno}
\begin{document}

\title[ ]{Noise-induced dynamics and photon statistics in bimodal quantum-dot micropillar lasers}

\author{Yanqiang Guo$^{1,2,*}$, Jianfei Zhang$^{1,2}$, Xiaomin Guo$^{1}$, Stephan Reitzenstein$^{3,*}$ and Liantuan Xiao$^{1,*}$}

\address{$^{1}$Key Laboratory of Advanced Transducers and Intelligent Control System, Ministry of Education, College of Physics, Taiyuan University of Technology, Taiyuan 030024, China\\
$^{2}$State Key Laboratory of Cryptology, Beijing 100878, China\\
$^{3}$Institut für Festkörperphysik, Technical University of Berlin, Hardenbergstr. 36, D-10623 Berlin, Germany
}
\ead{guoyanqiang@tyut.edu.cn, stephan.reitzenstein@physik.tu-berlin.de and xiaoliantuan@tyut.edu.cn
}
\vspace{10pt}
\begin{indented}
\item[]April 2023
\end{indented}

\begin{abstract}
Emission characteristics of quantum-dot micropillar lasers are located at the intersection of nanophotonics and nonlinear dynamics, which provides an ideal platform for studying the optical interface between classical and quantum systems. In this work, a noise-induced bimodal quantum-dot micropillar laser with orthogonal dual-mode outputs is modeled, and nonlinear dynamics, stochastic mode jumping and quantum statistics with the variation of stochastic noise intensity are investigated. Noise-induced effects lead to the emergence of two intensity bifurcation points for the strong and the weak mode, and the maximum output power of the strong mode becomes larger as the noise intensity increases. The anti-correlation of the two modes reaches the maximum at the second intensity bifurcation point. The dual-mode stochastic jumping frequency and effective bandwidth can exceed $100$ GHz and $30$ GHz under the noise-induced effect. Moreover, the noise-induced photon correlations of both modes simultaneously exhibit super-thermal bunching effects ($g^{(2)}(0)>2$) in the low injection current region. The $g^{(2)}(0)$-value of the strong mode can reach over 6 in the high injection current region. Photon bunching ($g^{(2)}(0)>1$) of both modes is observed over a wide range of noise intensities and injection currents. In the presence of the noise-induced effect, the photon number distribution of the strong or the weak mode is a mixture of Bose-Einstein and Poisson distributions. As the noise intensity increases, the photon number distribution of the strong mode is dominated by the Bose-Einstein distribution, and the proportion of the Poisson distribution is increased in the high injection current region, while that of the weak mode is reduced. Our results contribute to the development preparation of super-bunching quantum integrated light sources for improving the spatiotemporal resolution of quantum sensing measurements and enhancing the security of optical communication.
\end{abstract}

%
% Uncomment for keywords
%\vspace{2pc}

% Uncomment for Submitted to journal title message
%\submitto{\NJP}
%
% Uncomment if a separate title page is required
%\maketitle
%
% For two-column output uncomment the next line and choose [10pt] rather than [12pt] in the \documentclass declaration
%\ioptwocol
%

\section{Introduction}

The quantum-dot micropillar concept is a highly versatile and promising platform that allows for exciting research at the intersection of various fields, including nanophononics, nonlinear dynamics, and quantum optics \cite{Reitzenstein10,Christopher19,moody22,Weng18,Somaschi16,Ding16}.
Quantum-dot micropillar lasers (QDMLs) serve as cavity-enhanced devices and offer distinct advantages over conventional semiconductor lasers, including higher integration density for neuromorphic applications \cite{Heuser19}, lower threshold current \cite{Reitzenstein08}, and greater differential gain\cite{Kreinberg17}. In addition, bimodal QDMLs can output two orthogonal polarization modes induced by a slight elliptical asymmetry of the pillar’s cross section \cite{Heermeier22}. The small mode volume and ultra-high quality (Q) factor of quantum-dot micropillars lead to a significant Purcell effect, for instance to enhance the photon extraction efficiency of quantum light sources \cite{Ding16,Thomas21} and to increase the spontaneous emission coupling ($\beta$) factor of QDMLs \cite{Heermeier22}. Meanwhile, the mode-jumping dynamics in a high-$\beta$ bimodal QDML has been analyzed under weak initial noise and high injection current \cite{Redlich16}. This exploration holds significant importance in investigating the fundamental physical mechanisms of the QDML as a stochastic source.

The study of the QDML output properties involves both macroscopic dynamics and quantum statistics. The QDML dynamics indicate nonlinear input-output and spectral characteristics \cite{Reitzenstein11,Andreoli21}, dynamical mode splitting \cite{Reitzenstein10*,Gur21}, and polarization mode-switching \cite{Holzinger18,Leymann17}. Moreover, the rich and unique nonlinear dynamics of  QDMLs have been studied through external disturbances and coupling with active devices, such as frequency and phase locking with external injection \cite{Schlottmann16}, chaotic emission under optical feedback \cite{Albert11,Holzinger19}, and zero-lag synchronization by mutual coupling \cite{Kreinberg19}. In addition, research on the dynamic of noise-dominated effects focuses on delay-coupled semiconductor lasers \cite{Yoshimura07,Sunada12,GuoX20}, to study intriguing effects such as quantum noise injection for entropy enhancement and external period suppression of chaotic lasers \cite{GuoY21}, and optical phase synchronization driven by a  broadband noise \cite{Sunada14}. However, further insights into noise-induced dynamics and its effects on cavity-enhanced microlasers remain important, and it is important to explore so far unexposed issues, which are essential for advancing the field and unlocking the full potential of these systems for practical applications.

Meanwhile, photon statistics and correlations measurements are pivotal and fundamental for characterizing the quantum properties of QDMLs. The ground-breaking photon autocorrelation experiment conducted by Hanbury Brown and Twiss connected the temporal and spatial second-order degrees of coherence of starlight, marking a significant milestone in quantum optics \cite{HBT56,HBT56*}. The correlation function characterizing the coherence was formalized by Glauber in the 1960s \cite{Glauber63} and is used today to describe the nature of the photon emission process of light fields, which provides indispensable information on the photon statistics of the chaotic and quantum light sources \cite{GuoY22,GuoY18,GuoY18*,Lan17}. Furthermore, the quantum statistical analysis is important for the QDML applications of secure communication \cite{Paraiso21} and super-resolution imaging \cite{Israel17}, and photon correlation studies have been utilized to investigate its excitation spectrum \cite{Hopfmann13,Kazimierczuk15,Lettau18}, superradiant pulse emission \cite{Jahnke16,Kreinberg17}, and mode-switching effect \cite{Redlich16}. In addition, the photon number distribution of two-mode microlasers can be examined by a transition-edge sensor system to get access also to higher-order photon correlation functions \cite{Schmidt21,Schlottmann18}. The coupling of the QDMLs with external devices has allowed for the investigations of second-order coherence under optical feedback \cite{Holzinger18}, optical injection \cite{Schlottmann19,lingnau20}, and mutual coupling \cite{Kreinberg19}. However, the detailed study of noise-induced effects on the nonlinear dynamics QDMLs is still elusive and it is necessary to further explore the quantum statistics of the lasers with noise induction, which would lead to the development of new technologies such as quantum sensing and high-speed communication.

In this work we exploit a semiclassical four-variable rate
equation to model the dual-mode output of a bimodal QDML. We introduce a variable noise-induced intensity, which is presented as the Langevin noise in the slow-varying complex amplitude equation and includes a random noise intensity coefficient. With the stochastic noise-induced effect, the nonlinear dynamics and quantum statistics of the QDML are investigated. In this setting we obtain ultrahigh mode-jumping frequencies and wide effective bandwidths of the noise-induced QDML emission. Moreover, dual-mode parallel outputs of bunching and super-bunching light are available over a wide range of noise-induced output intensities and injection currents. The photon number distributions of the strong and the weak modes can be controlled by adjusting the noise-induced intensities and injection currents.

\section{Theoretical model}
\begin{figure}[htbp]
	\centering
	\includegraphics[width=1.0\textwidth]{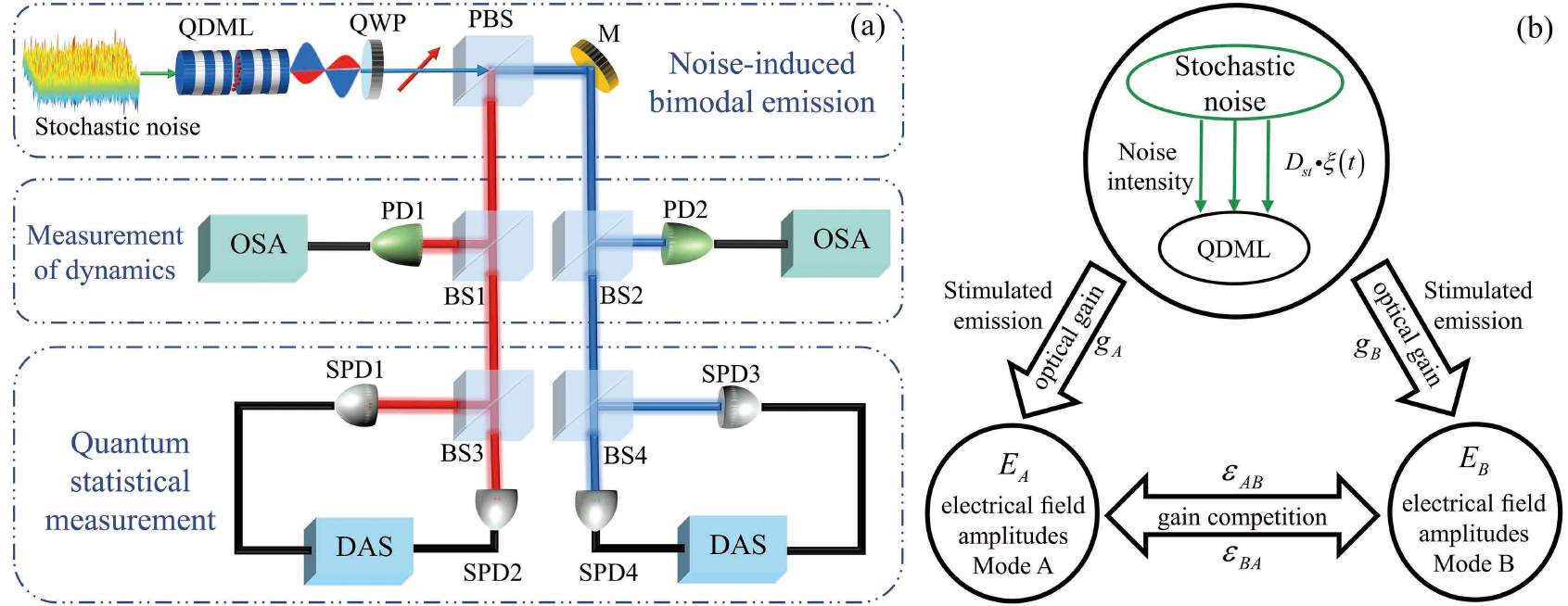}
	\caption{(a) Schematic diagram of dynamical and quantum statistical measurement for a noise-induced bimodal QDML. BS1 and BS2, are 10:90 beam splitters; BS3 and BS4, are 50:50 beam splitters; QDML: quantum-dot micropillar laser; QWP: quarter wave plate; PBS: polarized beam splitter; M: mirror; PD: photo detector; OSC: oscilloscope; SPD: single photon detector; DAS: data acquisition system. (b) Operating principle of noise-induced QDML.}
	\label{figure1}
\end{figure}

The schematic diagram for measuring the dynamics and quantum statistics of the noise-induced bimodal QDML is shown in figure \ref{figure1}(a). The output beam of the QDML is composed of two orthogonal linearly polarized modes and is parted through a quarter wave plate (QWP) and a polarized beam splitter (PBS). The strong mode (mode A) enters the 90:10 beam splitter (BS1), and $90\%$ of it is transformed into an electrical signal through a wide-bandwidth photodetector (PD1), which is recorded by a high-speed real-time oscilloscope and a spectrum analyzer for timing and spectrum dynamical measurements. The remaining $10\%$ output of the strong mode enters the beam splitter (BS3) and is separated into two beams of 50:50. The two beams are detected by two single-photon detectors (SPD1 and SPD2), and then the two electrical signals are processed by the data acquisition system (DAS) to obtain the second-order degree of coherence and photon statistical distribution of the strong mode . The weak mode (mode B) employs a similar scheme as mode A for measuring the dynamical and quantum statistical properties.

Figure \ref{figure1}(b) depicts the physical process of the QDML operation. The dual-mode output of the noise-induced QDML is subject to both stimulated radiation amplification and mode competition effects induced by gain coupling \cite{Redlich16}. A semiclassical four-variable rate equation is developed to model the output of the QDML. Furthermore, a Langevin noise term with a variable noise intensity coefficient is introduced into the slow-varying complex amplitude equation to mimic the noise-induced effect.
The equations of QDML with variable noise-induced intensities are described as follows:
\begin{eqnarray}
\frac{d}{dt} E_{j}(t)=\left[\frac{\hbar \omega}{\varepsilon_{0} \varepsilon_{b g}} \frac{Z_{Q D}^{a c t}}{V} g_{j}(2 \rho(t)-1)-\kappa_{j}\right](1+i \alpha) E_{j}(t)+D_{s t} \cdot \xi(t), \label{eq1}\\
\frac{d}{d t} \rho(t)=-\sum_{j \in\{A,B\}} g_{j}[2 \rho(t)-1]\left|E_{j}(t)\right|^{2}-\frac{\rho(t)}{\tau_{s p}}+S^{i n} n_{r}(t)[1-\rho(t)], \label{eq2} \\
\frac{d}{d t} n_{r}(t)=\frac{\eta}{e_{0} A_{e f f}}\left(I-I_{P}\right)-S^{i n} n_{r}(t) \frac{2 Z_{Q D}^{a c t}}{A_{e f f}}[1-\rho(t)] \nonumber\\
-S^{i n} \frac{2 Z_{Q D}^ {inac }}{A_{e f f}} \frac{\rho^ {inac }}{\tau_{s p}}-\frac{n_{r}(t)}{\tau_{r}},\label{eq3}
\end{eqnarray}
where $\xi(t)$ is a white Gaussian noise term and satisfies
\begin{eqnarray}
\left\langle\xi_{i}(t)\right\rangle=0,\left\langle\xi_{i}\left(t_{1}\right) \xi_{j}\left(t_{2}\right)\right\rangle=\delta\left(t_{1}-t_{2}\right),\label{eq4}
\end{eqnarray}
and $D_{s t}$ is the stochastic noise intensity. The subscripts $j \in\{A,B\}$ denote two orthogonal polarization modes respectively.  $E$ represents the slowly varying complex amplitude of the light field. $\rho$ represents the occupancy probability of active quantum dots. $n_{r}$ represents the reservoir carrier density. $\rho_{inac}(t)=\left(\tau_{s p} S^{i n} n_{r}\right) /\left(1+\tau_{s p} S^{i n} n_{r}\right)$ is the occupation probability of inactive quantum dots and the optical gain is $g_{j}=g_{j}^{0}\left(1+\varepsilon_{0} n_{b g} c_{0} \sum_{i \in\{A, B\}} \varepsilon_{j i}\left|E_{i}(t)\right|^{2}\right)^{-1}$.

The values for the QDML parameters used in the simulations are based on experimental data \cite{Kreinberg19} and are summarized in the following Table \ref{table1}.
\begin{table}
\caption{\label{math-tab2}QDML device parameters\cite{Kreinberg19} for the noise-induced QDML simulation .}
\begin{tabular*}{\textwidth}{@{}l*{15}{@{\extracolsep{0pt plus
12pt}}l}}
\br
Parameters&Symbol&Value\\
\mr
Optical cavity losses, mode A(B) &$\kappa_{A}\left(\kappa_{B}\right)$&$5.35(5.21)$ $ ns ^{-1}$ \\
Optical gain coeﬃcient, mode A(B)&$g_{A}^{0}\left(g_{B}^{0}\right)$&$5.35(5.21)$ $ m^{2} / \left(V^{2} s\right)$\\
Self-gain compression, mode A(B)&$\varepsilon_{A A}\left(\varepsilon_{B B}\right)$&$31(26) \cdot 10^{-10}$ $\mathrm{~m}^{2} / \left(A V\right)$\\
Cross gain compression, mode A(B)&$\varepsilon_{A B}\left(\varepsilon_{A B}\right)$&$31(42) \cdot 10^{-10} $ $\mathrm{~m}^{2} / \left(A V\right)$\\
Parasitic currents&$I_{p}$&$2.7 $ $\mu A $\\
Injection eﬃciency&$\eta$&$9.4 \cdot 10^{-2} $\\
Linewidth enhancement factor&$\alpha$&$1$\\
Reservoir carrier lifetime&$\tau_{r}$&$1$ $ns$\\
Purcell-reduced QD lifetime&$\tau_{s p}$&$230$ $ps$\\
\br
\end{tabular*}
\label{table1}
\end{table}

Due to the low mode volume of 4 $\mu m^{3}$ and a high Q-factor of $2\times10^4$, the QDML has significant light-matter interaction, which results in its optical field characteristics being strongly affected by stochastic noise.The noise-induced QDML source can be prepared by optical or electrical noise injection\cite{GuoY21}.The Purcell-enhanced emission reduces the QD lifetime to 230 ps, compared to a typical QD lifetime of 1 ns without the Purcell effect. To analyze the optimal output of the QDML and to verify the effect of noise-induced intensity on the QDML, we define the noise-induced intensity (NI) as the ratio of the stochastic noise intensity and the ground intensity at the laser threshold,
\begin{eqnarray}
N I(d B)=10 \log _{10}\left(\frac{D_{st }}{D_{tal }}\right)
\label{eq5}
\end{eqnarray}
where $D_{s t}$ is the stochastic noise intensity and $D_{tal }=\left|E_{A}^{t h}\right|^{2}+\left|E_{B}^{t h}\right|^{2}$ is the total intensity of the two modes at the laser threshold without noise induction. Accordingly, the electric field strength contributed by each photon is $E_{0}=\sqrt{(\hbar \omega) /\left(2 \varepsilon_{0} \varepsilon_{b g} V\right)}$, and the output power of the QDML can be expressed as $P_{j}=\hbar \omega \eta_{d} \kappa_{j}\left\langle\hat{n}_{j}\right\rangle^{2}$, considering the emission efficiency of the laser $\eta_{d}$ and the loss of the optical cavity $\kappa_{j}$, where $\hat{n}_{j}$ is the mean photon number of the laser output. Using this model we can study the dynamical and quantum statistical properties of the QDML under the influence of stochastic noise intensity.

\section{Results}
\subsection{Input-Output characteristics of noise-induced QDML}
\begin{figure}[htbp]
	\centering
	\includegraphics[width=1.0\textwidth]{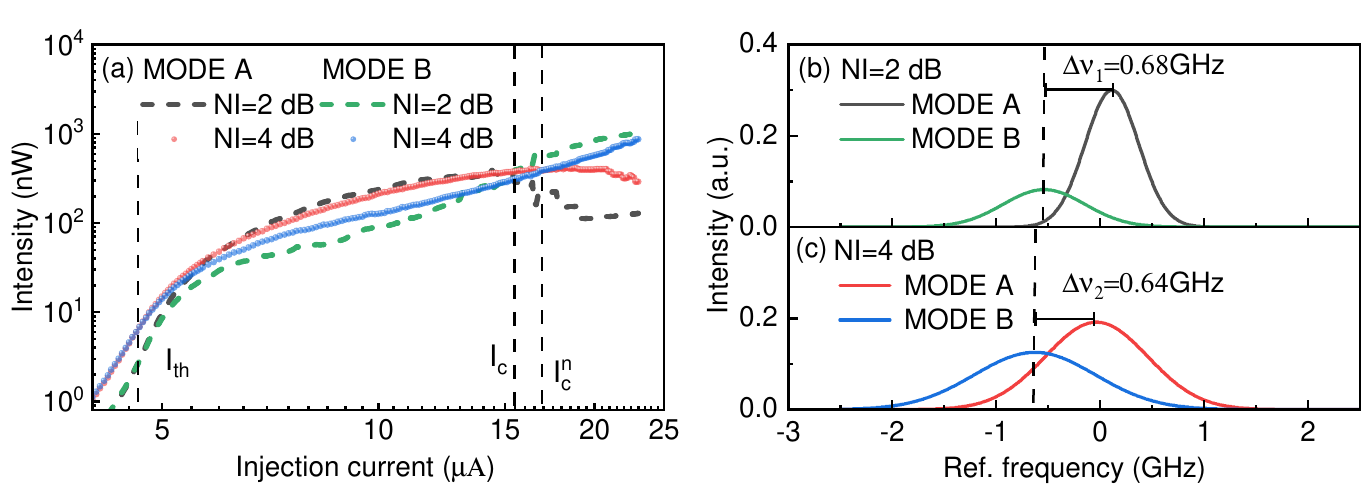}
	\caption{(a) Input-output characteristics of mode A and mode B of the QDML with different noise-induced intensities. The vertical lines denote the threshold current $I_{th}$, the injection current of the second intensity crossing point $I_{c}$ when $NI = 2$ dB, and the current of the second intensity crossing point $I_{c}^{n}$ when $NI = 4$ dB. (b)-(c) Optical spectra of mode A and B with different noise intensities when the injection current is $7$ $\mu $A. $\Delta\nu$ is the frequency difference of mode splitting.}
	\label{figure2}
\end{figure}

We first investigate the input-output characteristics and optical spectrum of the noise-induced QDML. Figure \ref{figure2}(a) shows the input current and output intensity characteristics of the QDML for different noise-induced intensities (according to equation (\ref{eq5}) $NI = 2$ dB and $NI = 4$ dB). In the low injection-current region, the output intensities of the strong and the weak mode are basically the same and grow linearly with an increase of the injection current. At the threshold current of $I_{t h}=4.6 $ $\mu \mathrm{A}$, the intensity of the two modes starts to increase nonlinearly, where the growth rate of mode A is higher than that of mode B, resulting in an intensity crossing point as observed also often in experiment \cite{Schmidt21}. As the injection current further increases, the output intensity of mode B continues to increase while mode A reaches saturation and decreases, leading to the second intensity crossing point of the two modes. In the QDML, the gain medium is limited to a finite number of quantum dots, resulting in competitive behavior between the two linearly polarized optical modes. At the threshold current, mode A prevails in the gain competition. As the injection current increases, mode A experiences gain saturation while mode B prevails in the gain competition and becomes the dominant mode.

Stochastic noise strongly affects the gain competition between the two modes, and the gain is redistributed as the noise-induced intensity increases, which leads to different output intensities of mode A and mode B for a given injection current. In the low injection region, the increase of noise-induced intensity increase the output intensity of the strong and weak modes but does not shift the threshold current point. In the medium injection current region, strong stochastic noise ($NI = 4$ dB) causes the overall output intensity of mode A to fall and that of mode B to rise visibly, compared to the case of weak stochastic noise ($NI = 2$ dB).  In the high injection current region, the output intensity of mode B decreases and that of mode A increases with a noise-induced intensity of $NI = 4$ dB, compared to the output with a noise intensity of $NI = 2$ dB. Moreover, the enhancement of noise-induced intensity contributes to shifting the second intensity crossing point towards a higher injection current. In fact, in figure \ref{figure2}(a), the injection current of the crossing point shifts from $I_{c}=15.6$ $ \mu \mathrm{A}$ (for $NI$ = $2$ dB) to $I_{c}^{n}=16.9 $ $\mu A$ (for $NI$ = $4$ dB).

Figures \ref{figure2}(b) and \ref{figure2}(c) illustrate the optical spectra of mode A and B for different noise-induced intensities while the injected current is held at $7$ $\mu $A. These results show that the fundamental mode is split due to a slight asymmetry of the elliptical cavity[10]. Upon further scrutiny, it becomes apparent that the increase of noise-induced intensity causes the mode splitting to decrease from 0.68 GHz to 0.64 GHz and broadens the spectra of both modes. The fluctuation of mode splitting implies a disturbance of stochastic noise induction on dual-mode gain competition, and the inhomogeneous spectrum broadening is due to the increase of output fluctuations as a result of the noise induction. The input-output findings provide insights into the complex interplay between stochastic noise and mode coupling and are beneficial for optimizing laser's output characteristics.

\subsection{Stochastic mode jumping and cross-correlation properties of noise-induced QDML}
\begin{figure}[htbp]
	\centering
	\includegraphics[width=1.0\textwidth]{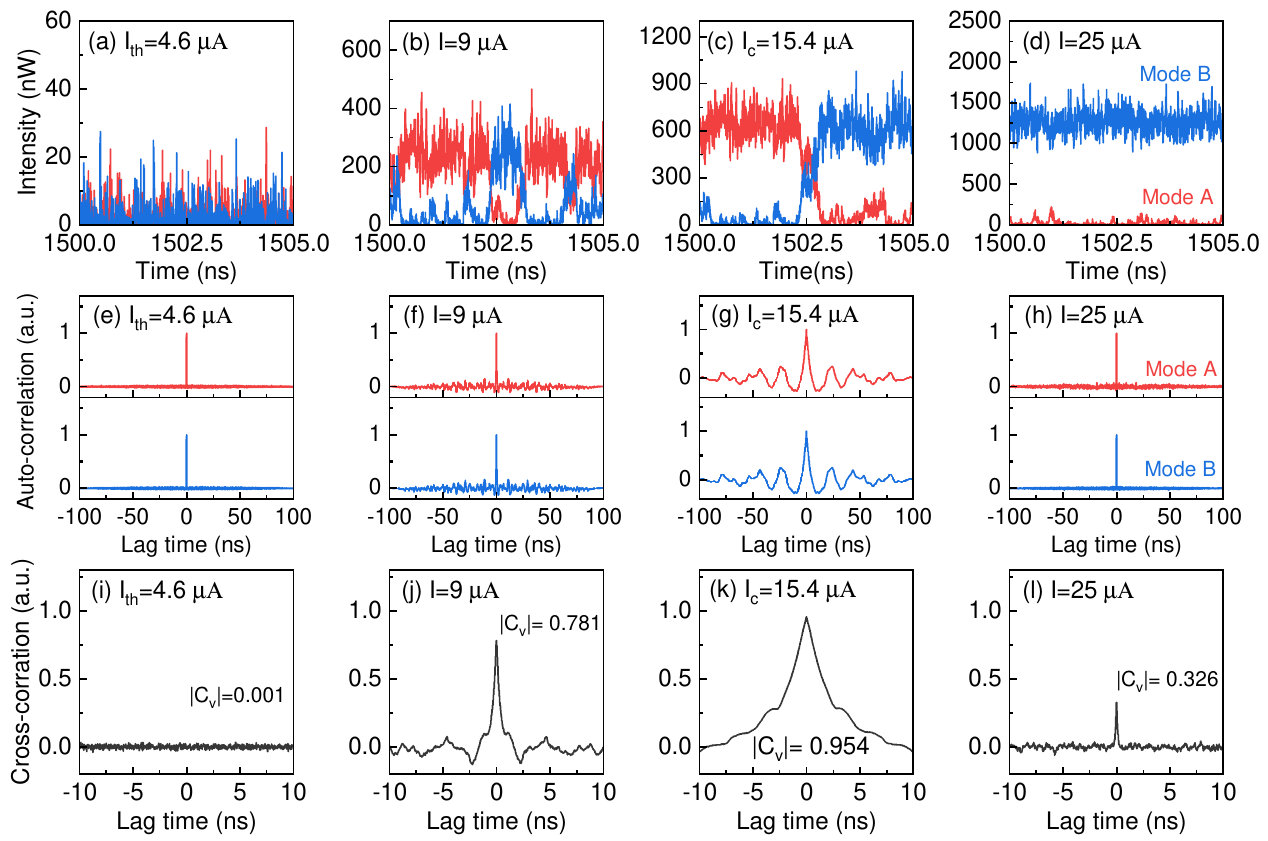}
	\caption{ Time series (a)-(d), auto-correlation(e)-(h) and cross-correlation (i)-(l) of the two modes, when the injection currents are $4.6 $ $\mu $A, $9 $ $\mu $A,$15.4 $ $\mu $A, and $25 $ $\mu $A with the noise-induced intensity $NI$ = $2$ dB. The red and blue lines represent mode A and mode B respectively. $\left|C_{v}\right|$ is absolute value of the cross-correlation at zero delay.}
	\label{figure3}
\end{figure}

Due to the gain competition between the strong and weak modes, the QDML exhibits distinct stochastic mode jumping behavior. In order to investigate the dynamical jumping between the two modes, we analyze the time series and correlation properties of the two modes for different injection currents with the noise-induced intensity $NI = 2$ dB. The results are shown in Figure. \ref{figure3}. The cross-correlation is defined as
\begin{eqnarray}
    C(\Delta T)=\frac{\left\langle\left[Int_{j}(t)-\left\langle Int_{j}\right\rangle \right]\left[Int_{j}(t+\Delta T)-\left\langle Int_{j}\right\rangle\right]\right\rangle}{\sqrt{\left\langle\left(Int_{j}(t)-\left\langle Int_{j}\right\rangle\right)^{2}\right\rangle\left\langle\left(Int_{j}(t)-\left\langle Int_{j}\right\rangle\right)^{2}\right\rangle}},\label{eq6}
\end{eqnarray}
where $Int$ is the output intensity of the laser,the subscripts $j \in\{A,B\}$ denote two orthogonal polarization modes respectively. $\Delta T$ is the delay time, and $\left\langle {}\right\rangle $ is the ensemble average. $C_{v}$ is the cross-correlation value at zero delay, and its absolute value $\left\vert C_{v}\right\vert $ is used to characterise the anti- correlation of the two modes. In addition, a mode jumping event is defined as a pulse of mode B with a higher intensity than that of mode A.

Figures \ref{figure3}(a), \ref{figure3}(e) and \ref{figure3}(i) show that at the threshold current of $I_{th}=4.6$ $\mu $A, the output intensity of the QDML is mainly contributed by the amplified noise, and there is strong auto-correlation and no cross-correlation between the two modes. In the effect of noise induction, the dual-mode anti-correlation becomes stronger as the injection current increases and mode jumping appears. At an injection current $I=15.4$ $\mu $A, i.e. at the intensity crossing point, the mode jumping frequency decreases and the auto-correlations of the two modes feature periodic oscillations, as shown in figures \ref{figure3}(c) and \ref{figure3}(g). These results indicate that the anticorrelated intensity fluctuations between mode A and mode B are most pronounced in the area of gain competition, and $\left|C_{v}\right|$ reaches a maximum value of 0.954 that exhibits the strongest anti-correlation as shown in figure \ref{figure3}(k). At the high injection current of $I=25$ $\mu $A, mode B becomes dominant winning the gain competition, and the mode jumping disappears. The anti-correlation between the two modes is also reduced, as shown in figures \ref{figure3}(d) and \ref{figure3}(l). And at this high current, the auto-correlations of both mode A and mode B exhibit high randomness as shown in figure \ref{figure3}(h), which indicates that the noise-induced QDML can simultaneously output two noise light fields for high injection currents.

\begin{figure}[htbp]
	\centering
	\includegraphics[width=1.0\textwidth]{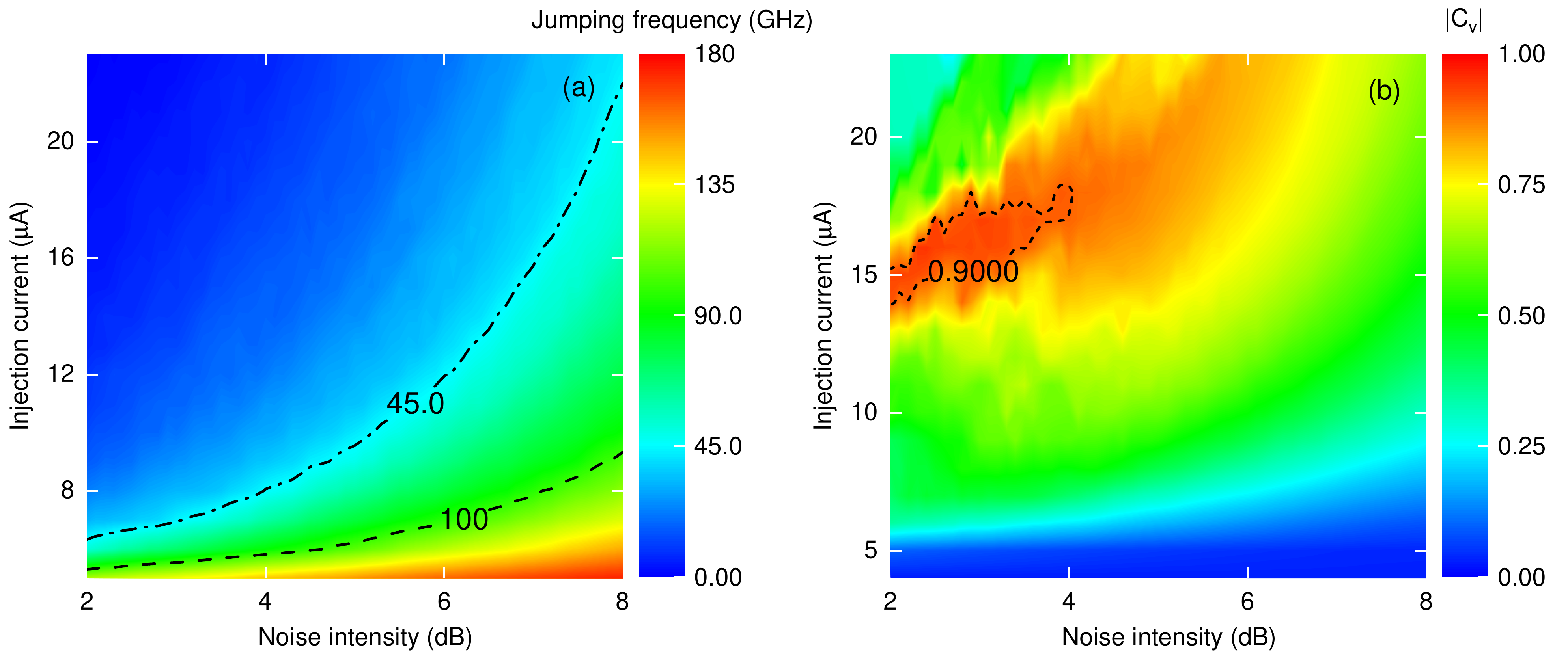}
	\caption{2D color maps of (a) mode jumping frequency and (b) cross-correlation versus noise intensity and injection current. The dashed-dotted and dashed lines in (a) represent the mode jumping frequencies of $45$ GHz and $100$ GHz respectively. The dashed region in (b) depicts the region of strong anti-correlation for $\left|C_{v}\right| \geq 0.9$.}
	\label{figure4}
\end{figure}

Figure \ref{figure4}(a) shows the map of the stochastic mode jumping frequency as functions of noise intensity and injection current. For one fixed injection current, the stochastic mode jumping frequency increases as the noise-induced intensity increases. With the fixed noise intensity, the mode jumping frequency decreases as the injection current increases. The dashed-dotted line indicates that the mode jumping frequency of $45 $ GHz. This high frequency can potentially allow for faster data transfer and more efficient data processing, compared to the conventional laser mode jumping frequency in the range of several gigahertz. As the noise intensity increases and the injection current decreases, the mode jumping frequency of the strong and weak modes can achieve even $100 $ GHz, corresponding to the dashed curve in figure \ref{figure4}(a). The ultrafast dual-mode jumping frequency implies the presence of fast alternating intensity random fluctuations between the dual modes \cite{Lettau18}, similar to random intensity fluctuations of feedback-coupled lasers, which is potentially useful for high-speed random number generation \cite{Kanter10}. Figure \ref{figure4}(b) shows the cross-correlation coefficient versus noise-induced intensity and injection current, and the dual-mode anti-correlation is characterized in terms of the absolute value of $C_v$. As the injection current increases, the anti-correlation of the two modes first increases and then decreases for a fixed noise-induced intensity. Moreover, as the noise-induced intensity increases, the anti-correlation of the two modes increases followed by a decrease for a fixed injection current above $10$ $\mu$A. The region marked by the dashed line represents a significant dual-mode anti-correlation ($\left|C_{v}\right| \geq 0.9$), which is around the intensity crossing point at $I_c=15.4$ $\mu$A. Strong correlations can be attained around low noise intensities and high injection currents. Importantly, these results indicate that the stochastic noise intensity boosts the gain competition between the two modes of the QDML and the dual-mode jumping. Furthermore, the notable anti-correlations between the two modes are revealed in the region near the second intensity crossing point. The anti-correlation increases as the noise intensity decreases, reaching the maximum $\left|C_{v}\right|= 0.954$ when the noise intensity is $2$ dB and the injection current is $15.4$ $\mu $A. The ultrafast mode jumping and strong dual-mode correlation are highly beneficial to optical synchronous communication and secure key distribution\cite{Martin16,ohara17}.

\subsection{Dual-mode effective bandwidth of noise-induced QDML}
To optimize the output frequency range of dual-mode random signals,we investigate the dual-mode effective bandwidth of the QDML under the influence of
\begin{figure}[htbp]
	\centering
	\includegraphics[width=0.85\textwidth]{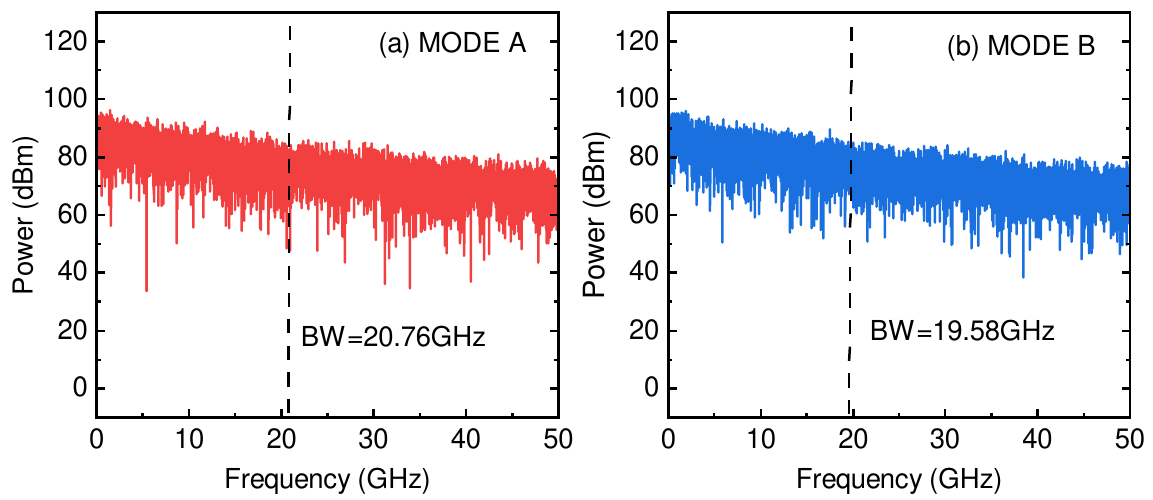}
	\caption{Frequency spectrum of (a) mode A and (b) mode B for $NI$ = $6$ dB and $I=10$ $\mu$A. The vertical lines denote the effective bandwidth of (a) mode A and (b) mode B.}
	\label{figure5}
\end{figure}
 noise-induced intensity. Figure \ref{figure5} shows the frequency spectra of the bimodal QDML for a noise intensity of $6$ dB and an injection current of $10$ $\mu$A. According to the definition of $80\%$ effective bandwidth, the effective bandwidth of the strong mode (mode A) is $20.76$ GHz, and that of the weak mode (mode B) is $19.58$ GHz.

\begin{figure}[htbp]
	\centering
	\includegraphics[width=\textwidth]{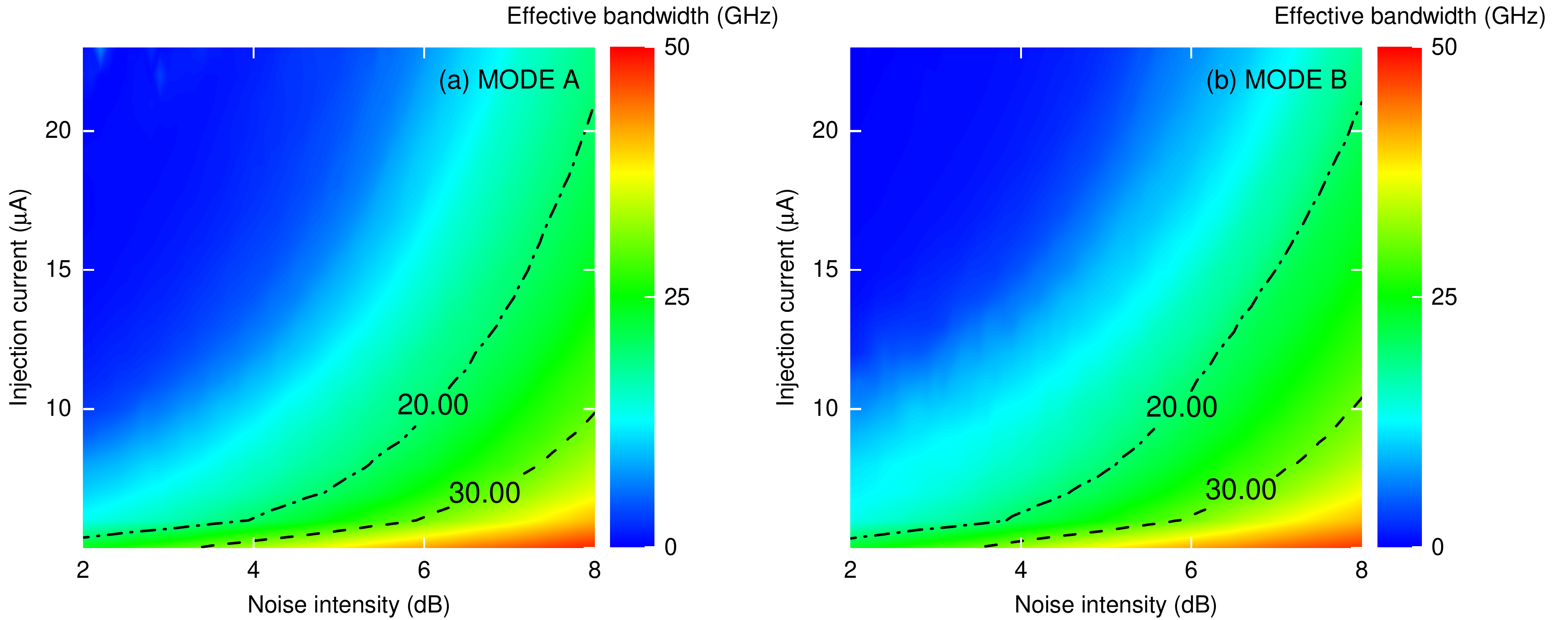}
	\caption{Effective bandwidth 2D color maps of (a) mode A and (b) mode B as functions of noise intensity and injection current. The dashed-dotted and dotted lines represent the effective bandwidths of $20$ GHz and $30 $ GHz for the two modes respectively.}
	\label{figure6}
\end{figure}

Figure \ref{figure6} shows the impact of noise intensity and injection current on the effective bandwidth of the bimodal QDML. When the injection current is held constant, the effective bandwidth increases as the noise intensity rises. On the other hand, when the noise intensity is held constant, the effective bandwidth decreases as the injection current increases. By adjusting the injection current and the noise intensity, the effective bandwidth of the bimodal QDML can be achieved above $20 $ GHz, and the boundary is indicated by the dashed-dotted line. Further increasing the noise intensity and decreasing the injection current, the effective bandwidth of the two modes can be simultaneously enhanced beyond $30$ GHz, as indicated by the region below the dotted limit in figure \ref{figure6}. It is worth noting that the increase of the stochastic noise intensity improves the randomness of the QDML output, thereby increasing the effective bandwidth.

\subsection{Photon correlation of the noise-induced bimodal QDML}
In addition to analyzing the dynamics of QDML, the quantum statistics is also important for revealing the coherence and probability statistical distribution of the laser output. The second-order photon correlation is an essential means to characterize the optical coherence and photon statistical behavior of the light field, which is defined as follows \cite{GuoY18*},
\begin{eqnarray}
    g^{(2)}(0)=\frac{\left\langle n_{1} n_{2}\right\rangle}{\left\langle n_{1}\right\rangle\left\langle n_{2}\right\rangle},\label{eq7}
\end{eqnarray}
Here, $n$ represents the photon number of the light field, and $\langle\rangle$ denotes the ensemble average. The $g^{(2)}(0)$ can reflect whether photons exhibit a tendency to arrive at detectors in pairs. If $g^{(2)}(0)$ is greater than 2, it indicates that the light is super-thermal light and exhibits photon super-bunching effect. If $g^{(2)}(0)$ is lower than the thermal limit of 2 and greater than the coherent limit of 1, it indicated the emission of thermal light with a photon bunching effect. And $g^{(2)}(0)$ below 1 implies a nonclassical photon anti-bunching.

\begin{figure}[htbp]
	\centering
	\includegraphics[width=0.85\textwidth]{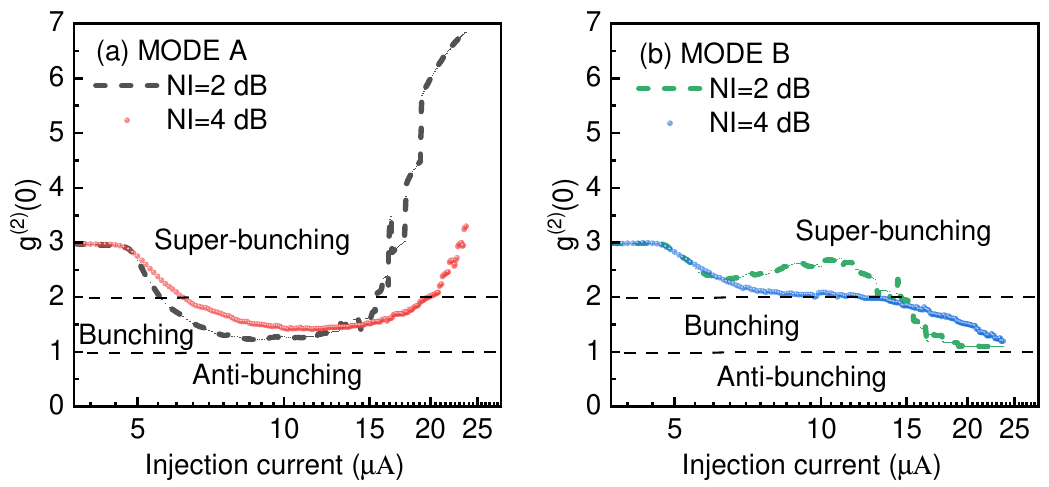}
	\caption{Second-order photon correlation of $g^{(2)}(0)$ versus injected current under the influence of noise induction. The dashed black and green lines represent mode A and mode B for $NI = 2$ dB. The red and blue highlights represent mode A and mode B for $NI = 4$ dB. $g^{(2)}(0)>2$ is the super-bunching region, $1<g^{(2)}(0)<2$ is the bunching region and $g^{(2)}(0)<1$ is the anti-bunching region.}
	\label{figure7}
\end{figure}

Figure \ref{figure7} shows the second-order photon correlation at zero time delay $g^{(2)}(0)$ of mode A and mode B versus the injection current under the influence of different noise intensities. At the threshold current $I_{th}=4.6$ $\mu $A (i.e., the first bifurcation point of $g^{(2)}(0)$), the $g^{(2)}(0)$ begins to decline from a value of $g^{(2)}(0)=3$, which indicates that the QDML output changes from super-bunching effect to bunching effect. When the injection current is $8.8$ $\mu $A and the noise intensity is $2 $ dB, the $g^{(2)}(0)$ of mode A falls to the minimum value of 1.22. Due to the gain competition, the $g^{(2)}(0)$ of mode B does not undergo a transition to the weak bunching state and instead remains in the super-bunching state ($g^{(2)}(0)\geq2$). As a result, the second-order photon correlation $g^{(2)}(0)$ of mode B reaches the maximum value of 2.70 when $NI = 2$ dB and $I =11 $ $\mu $A. At the second bifurcation point of the photon correlation, $g^{(2)}(0)$ of mode A begins to increase rapidly, while the $g^{(2)}(0)$ of mode B shows a downward trend. In the high injection current region, the mode A output transitions from a weak bunching state ($g^{(2)}(0)>1$) to a super-bunching state ($g^{(2)}(0)>2$), while mode B output changes to a weak bunching state. Furthermore, for $NI$ = $2$ dB and $I=20 $ $\mu $A, $g^{(2)}(0)$ of mode A exceeds 6 and the $g^{(2)}(0)$ of mode B is above 1. It is necessary to mention that the QDML without noise induction survives only one mode output which is a coherent state ($g^{(2)}(0)=1$).
This is attributed to the absence of noise-induced effect and only one mode dominates the gain.We conclude that in the present of noise induction the QDML can emit bunching states both in the strong mode and in the weak mode.

In figure \ref{figure7}, it is also verified that, for the same injection current, the noise-induced intensity has an inverse effect on the mode A and mode B. In the relatively low current region, the increase of noise-induced intensity enhances the photon bunching effect of mode A, and $g^{(2)}(0)$ of mode B decreases with increasing the noise intensity, leading to the mode B transition from the super-bunching state ($g^{(2)}(0)>2$) to the thermal state ($g^{(2)}(0)=2$ ). In the high current region, the bunching effect of mode A decreases as the noise intensity increases. For mode B, the increase of the noise intensity enhances its bunching effect, resulting in an increase of $g^{(2)}(0)$. The enhancement of the noise-induced intensity causes the $g^{(2)}(0)$ of the weak mode to remain around 2 over a wide range of injection currents. This is due to the increase of the noise-induced intensity enhancing the fluctuations of the optical field, which causes the weak mode to converge to a thermal state in the region of high injection currents. These findings suggest that the increase of the stochastic noise intensity leads to a shift in the dynamical regimes towards higher pump currents and, interestingly, the bunching behavior of the two modes can be continuously controlled by adjusting the noise intensity.

\begin{figure}[htbp]
	\centering
	\includegraphics[width=\textwidth]{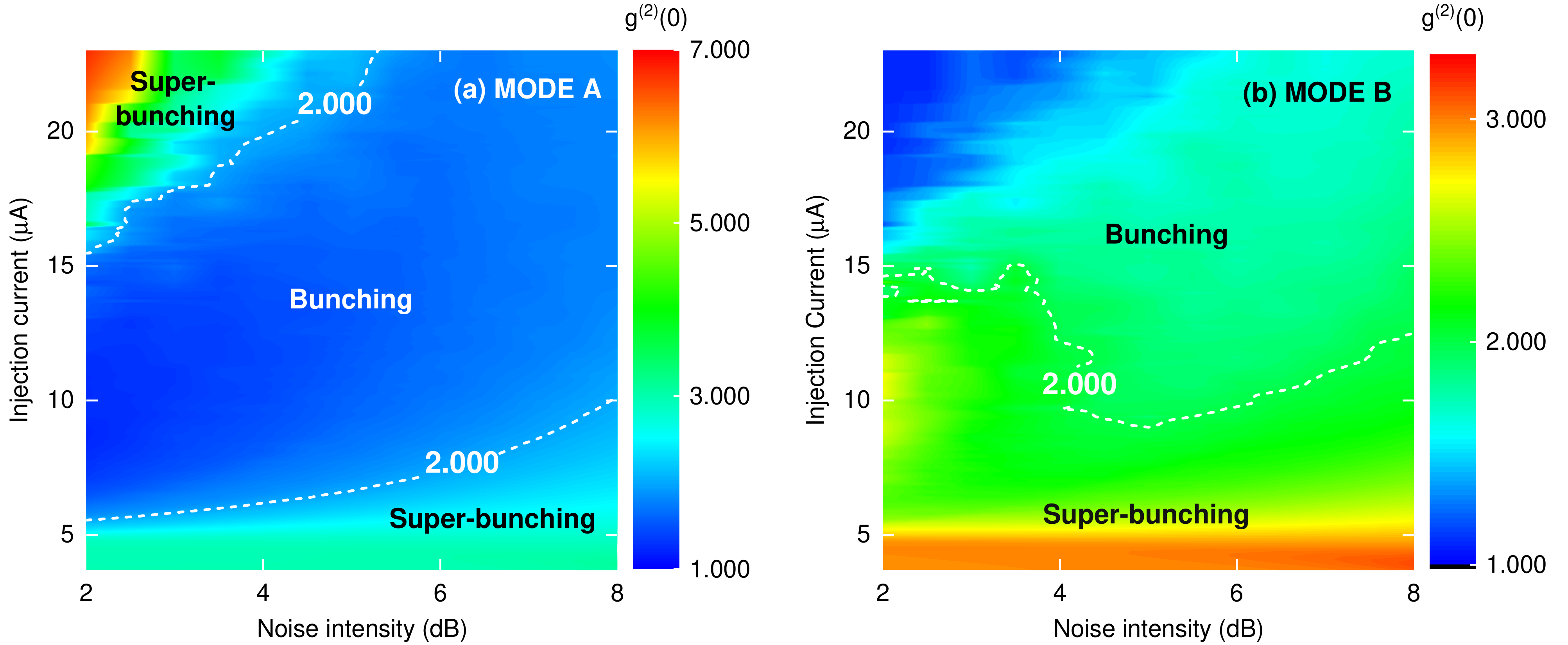}
	\caption{ Second-order photon correlation $g^{(2)}(0)$ of (a) mode A and (b) mode B as functions of noise intensity and injection current. The dotted line represents $g^{(2)}(0)=2$. $g^{(2)}(0)>2$ is the super-bunching region, $1<g^{(2)}(0)<2$ is the bunching region.}
	\label{figure8}
\end{figure}

To explore both the noise-induced and injection current effects on the photon correlation, we study $g^{(2)}(0)$ over a wide range of noise-current parameters. The results are shown in figure \ref{figure8} and indicate that the super-bunching states of mode A are present in the low current region and in the high current with low noise intensity region. The noise-current parameter region for the super-bunching outputs is $26\%$ of the overall operating region of figure \ref{figure8}(a). When the injection current is fixed and higher than $ 15$ $\mu $A, the photon correlation for the Mode A decreases as the noise intensity increases, but a higher photon correlation can be achieved by increasing the injection current, and it allows the output light with high power and strong bunching.
For mode B, the super-bunching states are mainly present in the low current region and account for $39\%$ of the total region of figure \ref{figure8}(b). In addition, the contour of $g^{(2)}(0)$ in the low injection current region shows complex variations. Stochastic noise boosts the mode competition and the evolution of dynamical characteristics\cite{Redlich16,lingnau20}, and for different noise-induced intensities the QDML photon correlation exists complicated bunching variations in different injection current regimes and low noise-induced intensities lead to unstable QDML outputs.It should be noted that the QDML can output emit light with bunching ($g^{(2)}(0)>1$) in parallel over a wide range of noise intensities and injection currents.

\subsection{Photon number distribution of the noise-induced bimodal QDML}
Finally, we investigate the photon number
distribution of the noise-induced QDML for various injection currents ($I =4.6$ $\mu $A, $I =15.4$ $\mu $A, $I =25 $ $\mu $A)
\begin{figure}[htbp]
	\centering
	\includegraphics[width=0.8\textwidth]{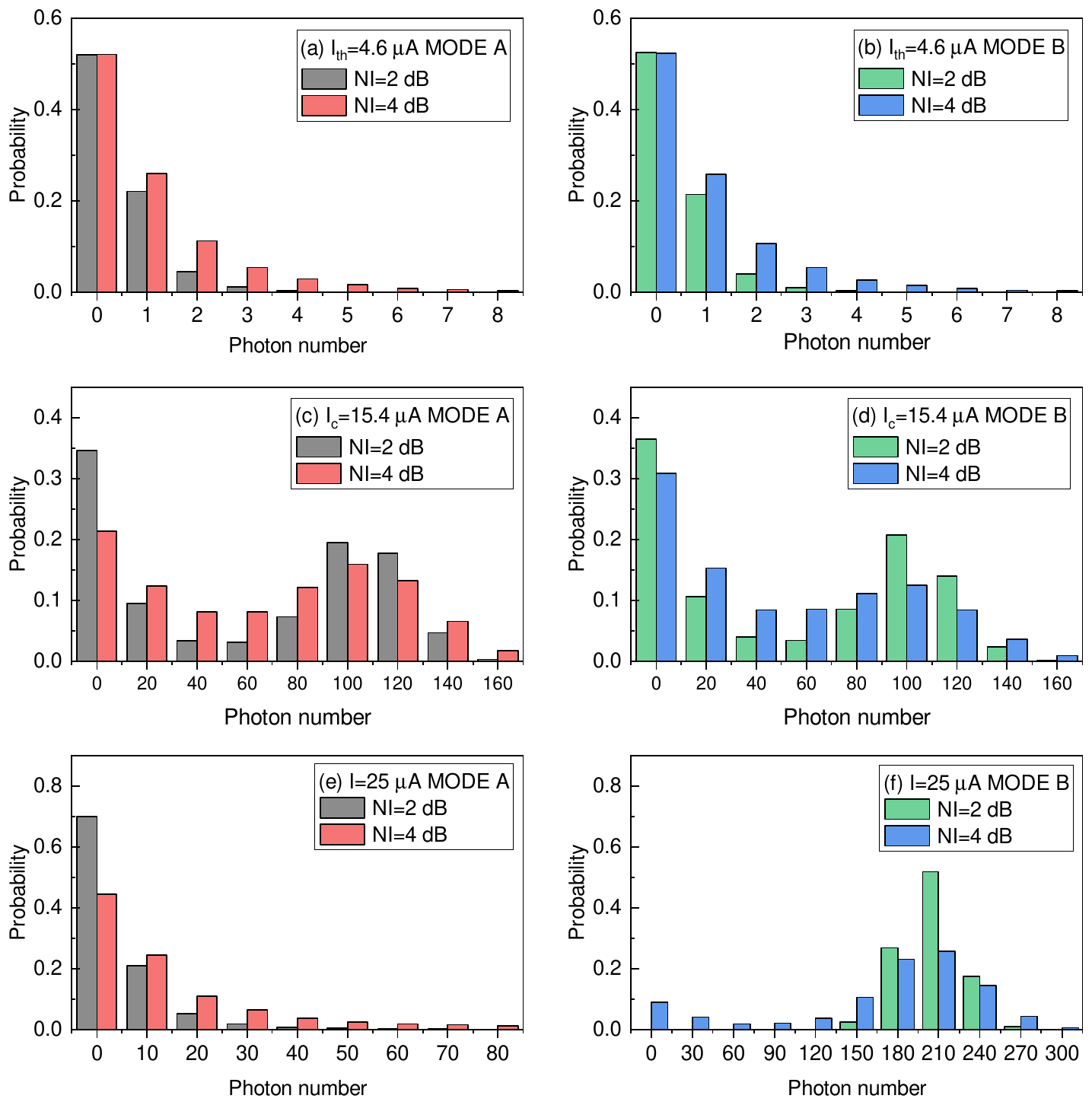}
	\caption{ Photon number distribution of noise-induced QDML for various injection currents of $4.6 $ $\mu $A, $15.4$ $\mu $A, and $25$ $\mu $A. The black and green bars represent the photon number distributions of mode A and mode B for $NI = 2$ dB. The red and blue bars represent the photon number distributions of mode A and mode B for $NI = 4$ dB.}
	\label{figure9}
\end{figure}
in figure \ref{figure9}. As the injection current is increased from the threshold value of $4.6$ $\mu $A to the intensity
crossing point of $15.4 $ $\mu $A, the photon number distribution of the bimodal QDML transitions from a Bose-Einstein distribution to a mixture of a Bose-Einstein and a Poisson distribution.In the high current region after the intensity crossing point, the proportion of the Bose-Einstein distribution in mode A increases, while the proportion of the Poisson distribution in mode B increases.

Figures \ref{figure9}(a)-\ref{figure9}(d) demonstrate the effect of noise intensity on the photon number distribution of mode A and mode B. As the noise intensity increases, the photon number distribution of both modes is consistently dominated by the Bose-Einstein distribution when the injection current is below $15.4$ $\mu $A (i.e., the second crossing point in figure \ref{figure7}). In this region, the increase of the noise intensity enhances the thermal emission of mode A, which increases the occurrence probability of multiphoton events. At the same time, the proportion of the super thermal emission of mode B decreases as the noise intensity increases. Figures \ref{figure9}(e) and \ref{figure9}(f) shows that the increase of noise intensity has a different effect on mode A and mode B above the injection current of $15.4$ $\mu $A. At a high current of $25$ $\mu $A, the increase of noise intensity decreases the thermal emission of mode A and increases the thermal emission of mode B, and the photon number distribution of mode A exhibits a super-bunching Bose-Einstein distribution and that of mode B approaches the Poisson distribution.

\section{Conclusion}
We employed a semiclassical four-variable rate equation to study the quantum statistics, nonlinear dynamics, and stochastic mode jumping of the bimodal QDML under stochastic noise induction. The noise-induced emission effect results in the appearance of two intensity crossing points for the strong mode (mode A) and weak mode (mode B), and the maximum output power of the strong mode increases as the noise intensity increases. The anti-correlation coefficient between the two modes reaches the minimum value of -0.954 at the second intensity crossing point. With the noise-induced effect, the dual-mode stochastic jumping frequency and effective bandwidth can exceed $100$ GHz and $30$ GHz, respectively. The parallel output of dual-mode bunching beams can be achieved by adjusting the noise intensity and the injection current. We also investigated the effect of noise intensity on the photon number distribution and find that the photon number distribution of the strong mode (mode A) is dominated by the Bose-Einstein distribution as the noise intensity increases. In the high injection current region, the proportion of the Poisson distribution of mode A increases as the noise intensity increases. Moreover, the photon number distribution of the weak mode (mode B) is also a mixture of Bose-Einstein and Poisson distributions. In the high injection current region, the proportion of the Bose-Einstein distribution of mode B increases as the noise intensity increases. The study sheds light on the effect of stochastic noise intensity on the QDML, with implications for the development of super-bunching quantum integrated light sources. The dual-mode integrated laser is of great importance to improve measurement resolution of quantum sensing by super-bunching photon correlation and enhance information security using wideband stochastic source.

\ack
The work was supported by the National Key Research and Development Program of China (2022YFA1404201), the National Natural Science Foundation of China (62175176, 62075154, 61875147), the Key Research and Development Program of Shanxi Province (International Cooperation, 201903D421049), the Shanxi Scholarship Council of China (HGKY2019023) and the Scientific and Technological Innovation Programs of Higher Education Institutions in Shanxi (201802053, 2019L0131).

\section*{Data availability statement}
The data that support the findings of this study are available upon reasonable request from the authors.

\section*{References}

\bibliographystyle{vancouver}

\bibliography{ref}

\end{document}